\documentclass[a4paper]{iopconfser}
\usepackage{amsmath}
\usepackage{amssymb}

\usepackage{url} 
\usepackage[breaklinks=true, colorlinks=true, linkcolor=blue, urlcolor=blue, citecolor=blue]{hyperref} 

\begin{document}
\title{The metric from energy-momentum non-conservation:\\ Generalizing Noether and completing spectral geometry}

\author{Achim Kempf}

\affil{Departments of Applied Mathematics and Physics, and Institute for Quantum Computing, University of Waterloo, Canada}
\affil{Perimeter Institute for Theoretical Physics, Waterloo, Canada}

\email{akempf@uwaterloo.ca}

\begin{abstract}
We complete the program of spectral geometry, in the sense that we show that a manifold's shape, i.e., its metric, can be reconstructed from its resonant sound when tapped lightly, i.e., from its spectrum, -- if in addition we also record the resonances' mutual excitation pattern when the driving is strong enough to reach the  nonlinear regime.
Applied to spacetime, this finding yields a generalization of Noether's theorem: the specific pattern of energy-momentum non-conservation on a generic curved spacetime, encoded within the quantum field theoretic scattering matrices, is sufficient to calculate the metric. 
Applied to quantum gravity, this shows that the conventional dichotomy of spacetime versus matter can emerge from an underlying information-theoretic framework of only one type of quantity: abstract correlators, $G^{(n)}$, that are, a priori, merely operators on $n$ tensor factors of a Hilbert space. This is because, on one hand, if abstract higher-order correlators $G^{(n>2)}$ can be diagonalized, these correlators can be represented as local quantum field theoretic vertices on a curved spacetime whose metric $g_{\mu\nu}(x)$ can be explicitly calculated.
On the other hand, at sufficiently high energies, such as the Planck scale, the $G^{(n)}$ may not be even approximately representable as correlators of a local QFT on a spacetime, indicating a regime that is mathematically controlled but transcends the concepts of spacetime and matter.
\end{abstract}

\section{Introduction}
$$$$
The conventional view is of a dichotomy of a spacetime which serves as a dynamical stage and matter fields that act on this stage. 
We here explore a new perspective suggested by the observation that, with increasing spacetime distance, the vacuum fluctuations in quantum fields become less correlated, which implies that a two-point correlator such as the Feynman propagator contains all information about distances. 
This observation was shown to allow replacing the notion of distance with the notion of correlation strength \cite{per}. 

Therefore, quantum field theoretic correlators describe not only the dynamics of matter, but also the underlying spacetime metric. This is hinting that the conventional dichotomy of spacetime and matter might be emerging from an underlying theory of only one type of quantity: correlations. In this picture, the notions of spacetime and matter emerge at low energies as approximate  representations of a pre-geometric theory of abstract correlators, see \cite{kempfpaper,reitzprl}. 

Our investigation is naturally closely related to the field of spectral geometry, the discipline which combines the mathematics of general relativity, i.e., differential geometry, to the mathematics of quantum theory, i.e., functional analysis, by asking `Can one hear the shape of a vibrating object?'. So far, the answer has been no. We show that one can, if one listens not only to the resonance frequencies, i.e., the spectrum, but also to the mutual excitations among the resonances when the amplitudes reach the nonlinear regime.

Technically, spectral geometry asks to what extent the geometry of a manifold is determined by the spectrum of a wave operator defined upon it \cite{kac, weyl, datchev}. It is known, however, that using the spectrum of the wave operator associated with $G_F$ (or $G^{(2)}$) alone has limitations. Spectra are often continuous for Lorentzian signature operators, and even when discretized as in the case of compact Riemannian manifolds, the spectrum represents invariants under the full unitary group, which is larger than the diffeomorphism group, thus typically providing incomplete geometric information \cite{kempfpaper, bhamre}. 

The spectrum of the wave operator, or its inverse, the 2-point function, therefore does generally not contain enough information to infer the metric. However, if we were given not only the spectrum of the wave operator or the 2-point function, i.e., if we were given them not only in their eigenbasis, but if were also given a position basis, then we could write the wave operator or 2-point correlator in position space. And the 2-point correlator, the Feynman propagator $G_F^{(2)}(x,y)$ in a position representation indeed contains sufficient information to reconstruct the metric tensor $g_{\mu\nu}(x)$ \cite{kempfpaper, per}. But then the question arises how, in addition to the spectrum of the wave operator or the propagator, one may also find a position basis. 

For example, say we are to determine the non-symmetric generic shape of a thin metal sheet in the shape of a potato. If we tap the metal lightly we can hear the resonance frequencies which tell us the spectrum of the Laplacian, and its inverse, on the manifold, which is not enough to infer the shape. How can we access the information needed to infer a position basis? If we can obtain a position basis then we can then write the Laplacian or its inverse in coordinates, from which we  can then calculate the metric. 

Here we will show that we can obtain a position basis and therefore infer the metric if we also excite the metal's oscillations strongly enough to enter the nonlinear regime, where the modes can excite other modes. Listening to this reveals the matrix elements of the interaction term - in QFT, this is essentially to measure the $S$-matrix elements in the eigenbasis of the wave operator. In this way, by listening to the anharmonicities of the potato's vibrations and thereby measuring the interaction term in the wave operator's eigenbasis, we now have a tool to find position bases: position bases are bases in which the interaction terms are singular at the coincidence of their arguments. At tree level, i.e., at the onset of nonlinearity, the interaction terms are even diagonal in position bases. To cover the cases of weak and strong nonlinearity simultaneopusly, we can here define the term 'diagonalizing' to possess a specific technical meaning: diagonalizing is here to seek a basis in which the interaction vertex $G^{(n>2)}$ exhibits the unique signature of locality — namely, it is singular if and only if all its arguments coincide. At tree level this corresponds to true diagonalization, while at loop level it corresponds to identifying this singular structure.  

We can also identify the present result as a generalization of Noether's theorem. To this end, let us recall that the absence of Killing vector fields and therefore the non-conservation of energy and momentum shows in interacting QFT on curved spacetimes in the vertices $G^{(n)}$ in the eigenbasis of the wave operator. Therefore, while the standard Noether theorem connects continuous symmetries to conserved quantities, our finding connects the specific way energy-momentum fails to be conserved in curved spacetime to the background spacetime geometry $g_{\mu\nu}$ that governs this non-conservation.

The structure of this paper is as follows: Section 2 briefly revisits the concept of replacing distance with correlation and the description of spacetime via the 2-point function. Section 3 discusses the standard spectral geometry approach and its limitations in this context. Section 4 presents the core new result: how interactions, through the principle of locality, allow for the completion of spectral geometry and the derivation of the metric from abstract correlators. Section 5 elaborates on the interpretation of this result as a generalized Noether theorem. Section 6 discusses the view of QFT on curved spacetime as an approximate representation of an underlying abstract structure of correlators. Section 7 offers concluding remarks and discusses future directions.

\section{Spacetime Geometry from the 2-Point Correlator}
$$$$
Let us begin by recalling the connection between spacetime geometry and quantum field correlators \cite{kempfpaper}. General relativity describes spacetime as a Lorentzian manifold $(M,g)$, where $M$ is a differentiable manifold and $g$ is the metric tensor. Equivalently, it can be described using the Synge world function $\sigma(x,x')$, representing half the squared geodesic distance between $x$ and $x'$ \cite{synge}. The metric can be recovered from $\sigma(x,x')$ via differentiation, requiring knowledge only near the diagonal $x=x'$ \cite{synge, fulling, birrelldavies}:
\begin{equation}
g_{\mu\nu}(x) = -\lim_{x\rightarrow x'} \frac{\partial}{\partial x^\mu}\frac{\partial}{\partial x'{}^\nu} \sigma(x,x')
\label{eq:metric_from_synge}
\end{equation}
This traditional geometric description relies implicitly on measurements using idealized rods and clocks to operationalize the notion of distance \cite{einstein}. However, such instruments are not fundamental, particularly at subatomic scales.

Our approach here replaces rods and clocks with the ubiquitous vacuum fluctuations of quantum fields. The correlations between these fluctuations naturally depend on the spacetime separation of the points considered. For instance, the Feynman propagator for a scalar field, $G_F(x,x') = \langle 0\vert T\hat{\phi}(x)\hat{\phi}(x')\vert 0\rangle$, exhibits a characteristic decay as the spacetime interval between $x$ and $x'$ increases, both into the timelike and spacelike regions \cite{kempfpaper, fulling, birrelldavies}. Inside the lightcone, this decay is due to the spreading of propagating perturbations; outside the lightcone, it reflects the decay of vacuum entanglement with spatial distance, stemming from the local coupling nature of field theories (e.g., via the Laplacian term in the Klein-Gordon equation) \cite{kempfpaper, valentini, reznik1, reznik2, sorkin, srednicki, jasonmaria}.

Crucially, the strength of this correlation acts as a proxy for spacetime distance. Just as the Synge function near the diagonal determines the metric, the Feynman propagator $G_F(x,x')$ near the diagonal also suffices, ultimately because the Feynman propagator obeys the Hadamard scaling behavior towards the coincidence limit, i.e., it behaves in a known way as a function of the geodesic distance. It was shown in \cite{per} that the metric tensor can be reconstructed directly from $G^{(2)}_F(x,x')$ (which we will often denote simply as $G^{(2)}(x,y)$ hereafter) via
\begin{equation}
g_{\mu\nu}(x) = -\frac{1}{2} \left(\frac{\Gamma(D/2-1)}{4\pi^{D/2}}\right)^\frac{2}{D-2} \lim_{x\rightarrow y} \frac{\partial}{\partial x^{\mu}}\frac{\partial}{\partial y^{\nu}} \left( G^{(2)}(x,y)^{\frac{2}{2-D}}\right)
\label{eq:metric_from_propagator}
\end{equation}
where $D$ is the spacetime dimension (with special handling for $D=2$). For intuition, notice that the divergence of the propagator on the lightcone indicates the lightcone directions everywhere and therefore determines the causal structure. The causal structure, however, is known to determine the metric up to conformal factor \cite{hawkingellis}. What is nontrivial is that the propagator's finite decay away from the lightcone also fixes that conformal factor. 

Therefore, the description of spacetime as $(M,g)$ can be replaced by the description as $(M, G^{(2)}(x,y))$, where $G^{(2)}(x,y)$ is the 2-point function of a suitable quantum field. This replaces the non-canonical measurement tools of rods and clocks with the intrinsic correlations of quantum fields. Mapping spacetime curvature could, in principle, be achieved by measuring these correlations, e.g., via quantum optical techniques \cite{ralph, milburn} or high-precision particle physics experiments \cite{kempfpaper}.

This formulation of curved spacetimes, while conceptually shifting focus from metric distance to correlation, still relies on the coordinates $x$ provided by a coordinate system the manifold $M$. To progress towards a fundamentally non-geometric description, we need to understand the information about the metric that is encoded in the 2-point correlator $G^{(2)}$ without presupposing that $G^{(2)}$ is given in a coordinate system. This leads us to consider the correlator $G^{(2)}$ as an abstract operator and investigate its only intrinsic properties, namely its spectrum, which leads us to spectreal geometry.  

\section{Spectral Geometry and its Limitations}
$$$$
If we consider the 2-point function $G^{(2)}$ not just as a function $G^{(2)}(x,y)$ but as the kernel of an abstract operator, namely its associated wave operator $W$, can we extract the metric from only the basis-independent properties of $G^{(2)}$, i.e., from its spectrum alone?  This brings us to the field of spectral geometry, famously encapsulated by Mark Kac's question: ``Can one hear the shape of a drum?" \cite{kac,yasamanmarco, weyl, datchev}. In the context of spacetime, this translates to asking: To what extent does the spectrum of a wave operator (like the d'Alembertian $\Box$ or Klein-Gordon operator $\Box + m^2$) defined on a Lorentzian manifold $(M,g)$ determine the metric $g$? 

Let $W$ be the wave operator whose Green's function (propagator) is $G^{(2)}$, satisfying $W G^{(2)} = \mathbb{I}$. If $W$ is self-adjoint with respect to a suitable inner product, it possesses a spectrum, $spec(W)$, which is a set of real numbers (eigenvalues). Since the spectrum is invariant under unitary transformations, which includes coordinate changes, i.e.,  diffeomorphisms, the elements of $spec(W)$ are geometric invariants. The question is whether this set of invariants is sufficient to reconstruct the geometry $g$.

For several reasons, the answer is generally no, especially for Lorentzian manifolds relevant to general relativity \cite{kempfpaper}:

1. Information Content of Spectra: For hyperbolic operators typical of Lorentzian signature spacetimes, spectra are often continuous and degenerate, making them inherently less informative than the discrete spectra often found in elliptic problems (like the standard Laplacian on a compact Riemannian manifold). While imposing boundary conditions (infrared cutoffs) can lead to discrete spectra, these spectra become heavily dependent on the chosen boundaries and conditions, obscuring the intrinsic geometric information.

2.  Invariance Group Mismatch: The spectrum $spec(W)$ is invariant under the action of the entire unitary group $U(\mathcal{H})$ on the Hilbert space $\mathcal{H}$ of field configurations. Diffeomorphisms form only a subgroup of $U(\mathcal{H})$. Consequently, the set of spectral invariants is generally smaller than the set of invariants under diffeomorphisms alone. There can exist non-isometric manifolds that are isospectral, i.e., that share the same spectrum for a given wave operator.

3.  Basis Information Loss: Fundamentally, knowing only the spectrum $spec(W) = \{\lambda_n\}$ is equivalent to knowing the operator $W$ (or its inverse) in its eigenbasis, where it takes a diagonal form $W_{diag} = \text{diag}(\lambda_1, \lambda_2, ...)$. However, the formula Eq. (\ref{eq:metric_from_propagator}) requires the propagator $G^{(2)}(x,y)$, which is the kernel of $W^{-1}$ in a position basis, i.e., in a coordinate system. To obtain $G^{(2)}(x,y)$ from $W_{diag}^{-1} = \text{diag}(1/\lambda_1, 1/\lambda_2, ...)$, one needs the unitary transformation $U$ that connects the eigenbasis of $W$ to a position basis. The spectrum itself does not provide this unitary transformation $U$.

We here only remark that an approach called infinitesimal spectral geometry was demonstrated in \cite{bhamre, kempfpaper} to partially overcome these issues by considering how infinitesimal changes in the spectrum relate to infinitesimal changes in the metric, a problem that is linear, solvable, and can be iterated to obtain finite changes of shape from finite changes of sound. This approach highlighted that for dimensions $D>2$, wave operators for scalar fields are insufficient because metric perturbations are tensorial and cannot always be faithfully expanded in a scalar eigenbasis, i.e., the linearization leads to matrix that is non-invertible because it is not a square matrix. This suggests that using wave operators for symmetric covariant $2$-tensor fields is necessary for infinitesimal spectral geometry \cite{bhamre} to work in arbitrary dimensions.

In the present paper, we seek a more direct and non-perturbative method to recover the full geometric information, via the crucial basis transformation $U$, from the ``sound" of the quantum fluctuations of a scalar quantum field.

\section{Completing Spectral Geometry by adding Interactions}
$$$$
The resolution to the missing basis information lies, perhaps surprisingly, in considering  quantum field theories that are interacting, rather than free ones. While the 2-point function $G^{(2)}$ and its associated spectrum capture the harmonic modes of the field, the interaction vertices, represented by the connected $n$-point correlation functions $G^{(n)}$ for $n > 2$, capture the anharmonicities, i.e., the non-linearities. It is precisely these interactions that encode the information needed to identify the position basis.

Let us remember that our aim is to consider the set of abstract $n$-point correlators $\{G^{(n)}\}_{n=2,3,4,...}$ as the fundamental description of our theory. These can be thought of as multi-linear operators acting on the Hilbert space $\mathcal{H}$ of the fields that we are summing over in the QFT path integral (or more accurately, on tensor products of $\mathcal{H}$ or its dual). For example, $G^{(2)}$ is essentially (up to considerations of the kernel) the inverse of the wave operator $W$. For a theory with a quartic interaction like $\lambda \phi^4$, the connected 4-point function $G^{(4)}$ represents the fundamental interaction vertex.

The key insight now relies on the principle of locality inherent in physical interactions within quantum field theories. In a position representation ${|x\rangle}$, this locality dictates a specific structure for the interaction vertices $G^{(n>2)}$. Namely, $G^{(n>2)}(x_1, ..., x_n)$ must exhibit singularities precisely when all spacetime arguments coincide, $x_1 = x_2 = ... = x_n$. For example, at tree level, the $\lambda \phi^4$ interaction vertex function $V^{(4)}$ involves delta functions enforcing this coincidence:
\begin{equation}
V^{(4)}(x_1, x_2, x_3, x_4) \propto \lambda \int d^4x \sqrt{|g(x)|} \delta(x_1-x) \delta(x_2-x) \delta(x_3-x) \delta(x_4-x)
\label{eq:vertex_position}
\end{equation}
In the field-space Dirac notation used here (representing the fields summed over in the path integral, not QFT states), this tree-level vertex is indeed diagonal:
\begin{equation}
\langle x_1, x_2 | V^{(4)} | x_3, x_4 \rangle \propto \delta(x_1-x_2)\delta(x_3-x_4)\delta(x_1-x_3) \times (\text{coupling factor})
\end{equation}
Beyond tree level, while loop corrections and renormalization mean the vertex is no longer strictly zero when arguments do not coincide, the signature of locality persists: $G^{(n>2)}$ remains singular if and only if all arguments coincide in a position basis.

This characteristic singularity structure provides the crucial link: Given the abstract operators $G^{(2)}$ and $G^{(n)}$ (for some $n>2$) in an arbitrary basis (e.g., the eigenbasis of $G^{(2)}$), we can identify a transformation $U$ to a position basis ${|x\rangle = U|k\rangle}$ by finding the $U$ that transforms $G^{(n)}$ into a form exhibiting singularities precisely at the coincidence of its arguments.

Conversely then, if we are given the abstract operators $G^{(2)}$ and $G^{(n)}$ (for some $n>2$, e.g., $n=4$) to arbitrary order in perturbation theory, in an arbitrary basis (say, in the eigenbasis of $G^{(2)}$ where $G^{(2)}$ is diagonal), we can identify any unitary transformation $U$ as a transformation to a position basis if it transforms the $n$-point correlator $G^{(n)}$ into a form where it is singular exactly at the coincidences of its arguments. We can then pick any such a unitary transformations $U$ to transform $G^{(2)}$ from its eigenbasis to a position bases, so that we can then calculate the metric from it. 
Therefore, the procedure to extract the metric from the abstract correlators $\{G^{(n)}\}$ is as follows \cite{reitzprl}:

\begin{enumerate}
    \item    Obtain Abstract Correlators:     Start with the abstract $n$-point functions $G^{(n)}$ for $n=2$ and at least one $n>2$ (e.g., $n=4$ for $\phi^4$ theory). These could be given, for instance, in the energy-momentum eigenbasis if the theory were defined on flat space, or more generally, in the eigenbasis of the wave operator $W$ associated with $G^{(2)}$. In the eigenbasis $\{|k\rangle\}$ of $W$, $G^{(2)}$ is diagonal: $(G^{(2)})_{k k'} = \delta_{k k'} / \lambda_k$. The vertex $G^{(4)}$ will generally be a non-diagonal tensor $(G^{(4)})_{k_1 k_2 k_3 k_4}$ in this basis.

\item   Identify Position Basis via Interaction Locality: Find a unitary transformation $U$ that reveals the local nature of the interaction vertex $G^{(n)}$ (for $n>2$). This means finding a unitary, $U$, that maps the eigenbasis of the wave operator, or $G^{(2)}$, to a new basis such that the transformed vertex $(G^{(n)}){x_1 ... x_n}$ in the new basis (obtained by applying $U$ to each index of $(G^{(n)}){k_1 ... k_n}$) exhibits singularities precisely at the coincidences of its arguments ($x_1=...=x_n$), characteristic of a local interaction when written in position space. The basis $\{|x\rangle\}$ defined by such a unitary change of basis $U$ can then be identified as a position basis. The existence of such a basis $U$ is contingent on the underlying abstract interaction being representable as local. If one position basis exists, others can as usual be generated by diffeomorphisms. At tree level, finding a position basis is of course regular diagonalization of a correlator $G^{(n)}$ in the strict sense that $(G^{(n)}){x_1 ... x_n}$) vanishes whenever the arguments do not coincide). At higher orders, the renormalized $G^{(n)}(x_1,...,x_n)$ is generally finite and nonzero when the arguments do not coincide, but for our purposes it suffices that it remains singular at the coincidence of the arguments.  

\item   Transform the Propagator:     Apply the same unitary transformation $U$ to the abstract propagator $G^{(2)}$ (which was diagonal in the $|k\rangle$ basis) to obtain its representation in the position basis.
To tree level, this yields the Feynman propagator as a function of spacetime points $x, y$. To higher orders, the renormalized propagator still obeys the Hadamard scaling towards the coincidence limit.  

\item      Calculate the Metric:     Use the position-space propagator $G^{(2)}(x,y)$ obtained in step 3 in the formula Eq. (\ref{eq:metric_from_propagator}) to calculate the metric tensor $g_{\mu\nu}(x)$.

\end{enumerate}
\noindent
This procedure demonstrates that the full geometric information of the spacetime manifold $(M,g)$ is indeed encoded in the set of abstract correlators $\{G^{(n)}\}$, provided the theory includes interactions ($n>2$) that are local. 

This also completes the program of spectral geometry, in the sense that it shows that one can ``hear the shape of the spacetime" (or of say a metal sheet in the form of a potato) if one excites fields on it strongly enough to reach the non-linear interaction regime. These anharmonicities reveal the underlying position space because the nonlinear interactions occur locally. The sound obtained from strong excitations includes not just the frequencies $\lambda_k$ from $G^{(2)}$ but also the mode-coupling information from $G^{(n)}$ for $n>2$.

\section{Generalizing Noether's Theorem}
$$$$
This connection between the properties of interactions and the properties of the spacetime metric can also be understood as a generalization of Noether's theorem. Noether's theorem, in its usual formulation, establishes a one-to-one correspondence between continuous symmetries and conservation laws. For example, in quantum field theory on Minkowski space, energy and momentum are conserved and correspondingly the Feynman rules for vertices include delta functions ensuring $\sum p_{in} = \sum p_{out}$. 

In contrast, on a curved spacetime with a generic metric, spacetime translational symmetry is broken, there does not exist a covariant Fourier transform and there are no plane waves that could serve as in- or outgoing energy-momentum eigenstates. Correspondingly, the Feynman rules for vertices do not contain energy-momentum conserving Dirac deltas. However, in a generalization of Noether's theorem, the specific pattern of non-conservation is in one-to-one correspondence to the exact metric of the space time.    

To see this, let us recall that in the context of our procedure for deriving the metric from correlators, we are determining the spectrum of $G^{(2)}$ and the matrix elements of ($G^{(n)}$) for $n>2$ in the eigenbasis of the propagator ($G^{(2)}$). In flat space, this eigenbasis is the energy-momentum basis $\{|p\rangle\}$ and the interaction vertex $G^{(n)}(p_1, ..., p_n)$ in this basis contains the delta function $\delta^{(D)}(\sum p_i)$ which describes energy-momentum conservation. 

On a curved background, the eigenmodes $|k\rangle$ of the wave operator $W$ are generally   not   plane waves, and the eigenvalues $\lambda_k$ are not simply related to $p^2$. When we express the interaction vertex $G^{(n)}$ in this basis $|k\rangle$, it will   not   contain a simple delta function enforcing conservation of the labels $k$. Instead, $(G^{(n)})_{k_1 ... k_n}$ will describe transitions where $\sum k_{in} \neq \sum k_{out}$ (schematically). The specific structure of these non-diagonal elements $(G^{(n)})_{k_1 ... k_n}$ reflects that energy and momentum fail to be conserved in the $|k\rangle$ basis due to the background curvature.

Our procedure (Section 4) then uses the information encoded in this specific pattern of non-conservation (i.e., the detailed structure of $(G^{(n)})_{k_1 ... k_n}$) to find the transformation $U$ to a position basis $\{|x\rangle\}$, i.e., to a basis in which the interaction is local, i.e., singular exactly at the coincidences of the arguments (and diagonal to tree level). Using $U$, we obtain the propagator $G^{(2)}(x,y)$ in a position basis and its functional form then fully reflects the curved geometry. We know this because the metric $g_{\mu\nu}(x)$ can be derived from $G^{(2)}(x,y)$.
Schematically, we have
$$
\mbox{Standard Noether:}~~~~~~~~~ \mbox{Conservation Laws}~ \leftrightarrow ~\mbox{Symmetries,}   
$$
while in the absence of spacetime symmetry, we have:
$$
\mbox{~~~~~Generalized Noether:}~~~~~~\mbox{Pattern of Non-conservation}~ \leftrightarrow~\mbox{Metric}
$$
In essence, an interaction operator $G^{(n)}$ acts as a probe. How it couples different free modes $|k\rangle$ reveals the underlying geometric structure. Measuring the scattering matrix elements (related to $G^{(n)}$) in the eigenbasis of the wave operator, (i.e., what is normally a basis of effectively asymptotically free states) effectively measures the specific way energy-momentum conservation is broken compared to flat space, and this detailed information is precisely what allows us to construct the metric $g_{\mu\nu}(x)$ responsible for it. 

\section{QFT on Curved Spacetime as a Representation}
$$$$
The framework developed here, where the metric $g_{\mu\nu}(x)$ is derived from a set of abstract $n$-point correlators $\{G^{(n)}\}$, naturally supports the view that the standard model of particle physics, being a QFT on a curved spacetime $(M,g)$, might be an   approximate representation   of a more fundamental, possibly non-geometric structure. Let us consider, therefore, the possibility that the underlying reality is described by a set of abstract algebraic objects $\{G^{(n)}\}$ (suppressing indices for different fields). 
The question then is whether this set of abstract $\{G^{(n)}\}$ can be represented mathematically, perhaps approximately, as a set of QFT correlation functions living on a spacetime manifold.

The procedure outlined in Section 4 provides the conditions for such a representation to exist: the interaction vertices $G^{(n>2)}$ must be at least approximately `diagonalizable', in the sense that bases can be found in which $G^{(n)}$ is singular exactly at coincidence of its argument. If they are, the `diagonalizing' basis can be interpreted as a position basis $\{|x\rangle\}$ on some manifold $M$: Transforming $G^{(2)}$ into this basis gives $G^{(2)}(x,y)$, from which a metric $g_{\mu\nu}(x)$ can be derived using Eq. (\ref{eq:metric_from_propagator}). The entire set $\{G^{(n)}(x_1, ..., x_n)\}$ that is obtained by transforming all abstract $G^{(n)}$ into this basis constitutes the QFT representation on the derived spacetime $(M,g)$. This perspective opens several possibilities:

\begin{itemize}
    \item 
     Emergence of Spacetime:     The existence of a spacetime description is not fundamental but   emergent. It emerges only in regimes where the abstract $n$-point correlators $G^{(n)}$ for $n>2$ are local, i.e., diagonalizable in the sense described above. For first results, see \cite{reitzprl}.

\item      Breakdown of Spacetime:     In regimes where the vertices $G^{(n>2)}$ are intrinsically non-diagonalizable in any basis (perhaps at the Planck scale), the abstract structure $\{G^{(n)}\}$ would   not   admit a representation as a local QFT on any classical spacetime manifold. This would signal the breakdown of the conventional dichotomy of spacetime and matter. The Planck scale might not be characterized by exotic phenomena of or on spacetime, but by the sheer inability of the fundamental entities, the $G^{(n)}$, to be represented   as correlators on a spacetime \cite{kempfpaper,reitzprl}. In such a pre-geometric and pre-matter regime, the $G^{(n)}$ are mere structure and may not even be interpretable as correlators, i.e., also their probabilistic interpretation may be merely a property that is emergent in the low energy regime. 

\item   Multiple Representations:     It is conceivable that the same abstract structure $\{G^{(n)}\}$ could admit different approximate representations in different regimes (e.g., different energy scales). These representations might even correspond to spacetimes of different dimensionality, see \cite{reitzprl}. The crucial factor determining the `perceived' dimension and geometry would be the structure of the interactions $G^{(n>2)}$ and whether they admit a diagonalization corresponding to a local theory in $D$ dimensions. This connects superficially to holographic ideas \cite{holoreview}, though the mechanism here is tied to the representability of local interactions rather than specific dualities.

\item   Information-Theoretic Foundation:  If the abstract $\{G^{(n)}\}$ are primary, physics might be fundamentally information-theoretic \cite{kempfpaper, mermin}. This is because correlators are at the heart of information theory, describing everything from probability distributions to mutual informations and multi-partite entanglement. 
For instance, the breakdown of the spacetime picture (non-diagonalizability of vertices) might relate to fundamental limits on information density or resolution, perhaps connected to concepts like bandlimitation and generalized Shannon sampling theory \cite{kempfpaper, shannon, ak-shannon1, ak-shannon2, ak-shannon3, ak-shannon4, ak-shannon5, ak-shannon6, ak-shannon7, ak-shannon8, ak-soda2}, which arise in models with a minimum length uncertainty \cite{ak-jmp, ak-ucr1, ak-ucr2, fabio} and which have also been applied, for example, to cosmology, see \cite{cosmo1}.
\end{itemize}
Overall, this perspective shifts the focus from assuming a spacetime and quantizing fields on it, to starting with abstract correlational structures and asking when and how notions like matter fields and a spacetime with dimension and metric can emerge as effective descriptions.

\section{Outlook}
$$$$
We are here considering the possibility that the underlying reality is described by a set of abstract algebraic objects $\{G^{(n)}\}$. We called them abstract correlators but, in principle, they need not always possess an interpretation in terms of correlations. They are at this point merely an abstract structure and the question arises what determines them, what consistency conditions, perhaps related to causality in some abstract sense they may obey. To answer these questions it is likely to be useful is to combine all of the $G^{(n)}$ into one object, as a formal power series:
\begin{equation}
W[J]:=\sum_{n=0}^\infty [G^{(n)}\vert J\rangle^{\otimes n}]
\end{equation} 
In regimes where the $G^{(n)}$ can be represented as the $n$-point correlators of a QFT on curved spacetime, this (or its exponential) is of course the partition function and its Fourier transform from $\vert J\rangle$ to the new variable $\vert \phi\rangle$ is $e^{iS[\phi]}$, with $S$ the classical action. Even in regimes where there is no representation in terms of quantum fields on a curved spacetime, this Fourier transform can be performed, at least as a Fourier transform of formal power series, see \cite{KMJ}, and the $G^{(n)}$ can be studied in terms of $S$.  There are several additional avenues to pursue, for example:
\begin{itemize}
    \item    Mathematical Formalism:     Developing the functional analysis tools to handle abstract $n$-point operators $G^{(n)}$ basis-independently, characterizing their diagonalizability properties.
\item        Information-Theoretic Analysis:     Exploring the structure of the abstract $\{G^{(n)}\}$ using tools from quantum information theory. Can all fundamental principles be formulated purely in terms of entities such as information content, (quantum) channel capacities and entanglement measures?
 \item       Quantum Reference Frames:     Relating this framework to existing work on quantum reference frames, where relational descriptions and correlations play a central role \cite{qrf, gia2, est}. Some of the `position bases' derived here could potentially be understood within a framework of relative localization defined by interactions.
       \item  Relation to other approaches quantum gravity:  Investigating links to existing quantum gravity approaches like Loop Quantum Gravity \cite{loops}, Causal Dynamical Triangulations \cite{loll}, indefinite causality \cite{ics} or causal fermion systems, \cite{cfs}. Can the abstract correlators $\{G^{(n)}\}$ serve as a bridge or common language?
\end{itemize}
In conclusion, by incorporating interactions into spectral geometry, one can not only make spectral geometry work and generalize Noether's theorem to obtain an equivalence between non-conservation and the metric. One also obtains a framework in which the conventional dichotomy of a spacetime and matter can emerge as the low energy limit of a pre-geometric theory in which all degrees of freedom are described in a unified way, namely in terms of abstract correlators. This opens a path towards perhaps understanding spacetime and matter as an emergent representation of a deeper, information-theoretic reality. 
$$$$
\bf \noindent Acknowledgments \rm \medskip
\newline\noindent
AK acknowledges support through the Dieter Schwarz Foundation, a Discovery Grant of the National Science and Engineering Council of Canada (NSERC) and a grant from the National Research Council of Canada (NRC). AK also appreciates fruitful discussions and collaborations that informed this work, particularly with B. \v{S}oda and M. Reitz. AK is grateful to the hospitality of Thomas Elze, the organizer of the DICE 2024 conference where the results in the present paper were first presented. The present work will appear in the proceedings of DICE2024.

\end{document}